# Bubble-Driven Flow Transitions in Evaporating Active Droplets on Structured Surfaces

Meneka Banik and Ranjini Bandyopadhyay*

Soft Condensed Matter Group, Raman Research Institute, C. V. Raman Avenue, Sadashivanagar, Bangalore 560080, India.

* Corresponding author

Email: ranjini@rri.res.in

**Abstract**

The evaporation of particle-laden droplets on engineered surfaces underpins a wide range of technologies, from printed electronics to biosensing. While the influence of substrate topography on passive particle deposition is well established, the combined effects of active matter dynamics, catalytic gas generation, and surface structuring remain unexplored. Here, we investigate the drying of aqueous droplets containing Janus particles (polystyrene-platinum, PS-Pt) on topographically patterned substrates in the presence of hydrogen peroxide ($H_2O_2$) fuel. The catalytic decomposition of $H_2O_2$ produces oxygen bubbles within the droplet, introducing strong, transient hydrodynamic perturbations that compete with evaporation driven capillary flows and contact line interactions. We show that bubble activity alters particle transport, leading to distinct and tunable final morphologies not achievable with passive suspensions. This study demonstrates how bubble-induced flow coupled with substrate topography determines deposition patterns during droplet evaporation. Our findings open a route to harnessing active matter and reaction driven flows for directed particle assembly.

**Introduction**

The drying of colloidal droplets is a paradigmatic example of nonequilibrium self-assembly, widely leveraged in surface coating, inkjet printing, biosensing, and microfabrication technologies (1-3). A differential evaporation flux drives capillary flows and convects particles towards the pinned contact line, resulting in highly non-uniform ring-like deposits well-documented in the literature (4-6). This so-called coffee-ring effect arises from the coupling of evaporation kinetics, interfacial tension gradients, and hydrodynamic flows within sessile droplets, first rigorously quantified by Deegan et al. (4). The practical implications of such flow-induced heterogeneities have motivated numerous attempts to suppress coffee-ring formation and to achieve uniform films essential for optical coatings, electronic devices, and biosensors (1-3,7-9). Despite extensive study, controlling particle deposition morphologies formed via suspension droplet drying remains challenging. Various strategies have been proposed, including adding surfactants or polymers to modify internal Marangoni flows (10-12), tuning particle shape, and interactions to influence assembly dynamics (13,14), and engineering the substrate properties to affect droplet wetting and evaporation behavior (15,16).

Surface topography alters droplet wetting behavior, contact line dynamics, and spatial evaporation flux distributions, thereby tuning internal flow fields and deposit morphologies. Structured substrates featuring micro-pillars, grooves, or curvature impose geometric constraints that reshape the evaporative flux profile away from the classical edge-dominated regime (17-20). For example, convex substrate curvature redistributes vapor flux to increase evaporation at the apex, while attenuating the outward capillary flow results in particle accumulation at the contact line and promoting more uniform deposition (21). Furthermore, controlled contact line pinning on rough or chemically heterogeneous surfaces can induce intermittent stick-slip motion, as elucidated in the models of Stauber et al. (22), resulting in spatially modulated deposits including multi-ring patterns and complex particle arrays (23). The interplay of capillary forces, contact angle hysteresis, and surface energy variations govern the dynamics of pinning and depinning events, influencing the flow structures within the evaporating droplet (24). Despite advances, understanding flow-structure interactions on complex patterned surfaces is still in a nascent stage, particularly in regimes where competing hydrodynamic and Marangoni stresses coexist. For example, the magnitude and directionality of Marangoni flows can suppress or reverse coffee-ring formation, and depend on surface-induced thermal and solutal gradients, substrate wettability, and topographic features (18). While much of the existing literature focuses on passive colloidal particles, recent advances in active matter have introduced new paradigms in particle assembly. Polystyrene-platinum (PS-Pt) Janus particles, which catalytically decompose hydrogen peroxide ($H_2O_2$) fuel to oxygen and water, represent a prominent example of self-propelled colloids (25). This reaction generates local gradients in reactant and product concentrations, producing self-diffusiophoretic slip velocities and active propulsion (26,27). Beyond diffusiophoresis, the nucleation and growth of oxygen microbubbles on or near the catalytic cap introduce bubble propulsion mechanisms that significantly enhance particle velocity and alter their trajectories (28). Bubble dynamics induce strong hydrodynamic perturbations, including localized vortices and transient levitation forces that decouple particles from the substrate (29,30). The detachment and collapse of bubbles generate time dependent flow fields and pressure gradients which couple to particle motion and influence local fluid mixing, particle aggregation, and deposit morphology (31). Moreover, the lifetime, size distribution, and frequency of bubble generation are highly sensitive to fuel concentration, particle surface properties, and ambient conditions (32). In confinement, the interplay of bubble nucleation with interfacial evaporation and contact line dynamics leads to complex phenomena that challenge existing theoretical models developed primarily for bulk or quiescent fluids (33).

The intricate coupling of evaporation driven hydrodynamics, active propulsion, bubble-induced flows, and substrate topography defines an emergent multiphysical system with nonlinear, time dependent behavior largely unexplored in the available literature (34,35). Surface patterning modulates wettability and local curvature, influencing nucleation sites and stability of catalytic oxygen bubbles (36). This can alter bubble lifetime and size distribution, which in turn affect propulsion forces and particle-surface interactions. Simultaneously, evaporation-induced capillary flows interact with bubble-generated local jets and vortices, which should result in complex flow superpositions and particle transport pathways. Marangoni stresses, arising from thermal or solutal surface tension gradients modulated by substrate properties and particle activity, further complicate internal flows by generating recirculating convection patterns that compete with or reinforce capillary driven transport (18). The transient stick-slip motion of the contact line on patterned substrates dynamically modulates the droplet footprint and flow fields, thereby affecting active particle trajectories and bubble dynamics in a coupled fashion (37). Most literature focuses either on passive particle evaporation on patterned or functionalized surfaces (38), or catalytic microswimmer propulsion in bulk fluid environments without evaporation (39). The integrated effects of surface topography, evaporative transport, active propulsion, and bubble-mediated hydrodynamics remain a critical gap, limiting predictive control over final particle deposition and self-assembly outcomes in applied systems.

In this work, we systematically investigate the evaporation-driven drying of droplets containing PS-Pt Janus particles on patterned substrates under catalytic $H_2O_2$ decomposition conditions. We aim to elucidate how oxygen bubble pinning onto the substrates and evaporation-induced hydrodynamics influence particle transport, collective behavior, and final deposit morphologies. While bubble-mediated flow transitions (38) have been reported in microfluidic environments, ours is the first rigorous study wherein we evaluate the effect of bubble-substrate interactions. This study addresses a fundamental knowledge gap at the intersection of active matter physics, interfacial fluid mechanics, and surface science. A detailed mechanistic understanding of these coupled processes will enable design principles for dynamic, energy-consuming colloidal assemblies and advanced strategies for fabricating functional surfaces with tunable structures driven by internally generated chemical and hydrodynamic fields. The outcomes have broad implications in microfabrication, biomedical diagnostics, and active materials engineering.

**Materials and Methods**

**Synthesis of Janus particles**. Polystyrene (PS) microspheres with a mean diameter of 2.07 ± 0.15 µm (Bangs Laboratories Inc.) were used for Janus particle fabrication. A suspension of PS microspheres was diluted in isopropyl alcohol (IPA) to a final concentration of 0.5 wt %. Clean glass substrates were prepared by sequential ultrasonication in acetone, ethanol, and Milli-Q water (15 min each), followed by treatment in piranha solution (3:1 mixture of $H_2SO_4$ and $H_2O_2$). The substrates were rinsed thoroughly with Milli-Q water, dried in a hot air oven, and purged with nitrogen before use. A close packed monolayer of PS microspheres was obtained on the cleaned glass slides by the sedimentation method, producing a monolayer close packed array of particles over large areas (Fig. 1(a), left SEM micrograph). A 20 nm thick platinum (Pt) film was then deposited onto the exposed hemisphere of the PS particles using a Benchtop sputter coater (Quorum Q150RS). The Pt coating was confirmed by field emission scanning electron microscopy (FESEM), which revealed half coated PS particles (Fig. 1(a), middle SEM micrograph). The Janus particles were released from the substrate into Milli-Q water via ultrasonication, yielding a stock suspension. This stock was subsequently diluted to prepare 1 wt % aqueous suspensions of Janus particles (40,41). To induce self-propulsion, hydrogen peroxide ($H_2O_2$, 30 wt %, Thermo Fisher Scientific) was added to the suspensions to achieve final concentrations between 0.1 and 10 wt % (26).

**Fabrication of topographically patterned substrates**. Topographically patterned substrates were fabricated by soft lithography using Sylgard 184 (a two-part crosslinkable polydimethylsiloxane (PDMS) elastomer; Dow Corning, USA) films. The oligomer (part A) to cross-linker (part B) ratio was maintained at 10:1 (wt/wt). The Sylgard mixture was poured on two types of patterned templates (i) a student optical grating, OG (Hilger Analytical) with stripe height $h_P$ = 1 µm and line width $l_P$ = 5 µm, and (ii) a commercially available compact disc, CD (Verbatim) with stripe height $h_P$ = 120 nm and line width $l_P$ = 750 nm. Next, the Sylgard films were cured at 120 °C for 12 hours in a hot air oven for complete crosslinking and pattern replication, after which they were carefully peeled from the template, generating a negative replica of the original template, OG PDMS and CD PDMS respectively (42). Atomic Force Microscopy (AFM, Oxford Instruments MFP-3D-bio-AFM in tapping mode, with resonant frequency of 157 kHz) images of the replicated structures on Sylgard are shown in Fig. 1b. Some of the OG PDMS substrates were exposed to oxygen plasma treatment for 2 minutes (Plasma Prep system, SPI Supplies). Plasma irradiation removes surface methyl groups (-Si-$CH_3$) from the native PDMS, leading to the formation of silanol (-Si-OH) groups and an oxidized silica-like ($SiO_x$) surface layer. This chemical modification drastically reduces the hydrophobicity of Sylgard and transforms it to a highly hydrophilic state. The plasma-induced silanol groups also increase surface energy, thereby enhancing particle-substrate adhesion.

**Droplet Drying**. All droplet evaporation experiments were conducted under controlled ambient conditions of 23 °C and 40 % relative humidity. 2 µL suspension droplets were carefully dispensed onto pre-cleaned substrates using a calibrated micropipette (Tarson). The drying dynamics were monitored in-situ using a stereo zoom microscope (Olympus SZX16). Simultaneously, the evolution of the droplet contact angle was recorded using a contact angle goniometer in sessile droplet mode (Attension Theta Flex- Biolin Scientific Optical Tensiometer), enabling quantitative analysis of wetting behavior and evaporation kinetics. Droplet images were acquired at a frame rate of 2.3 fps, and were analyzed using the OneAttension software. After complete evaporation, the final dried morphologies were characterized using an FESEM. Optical microscope movies of the drying process were recorded at a frame rate of 3 frames per second. After acquisition, particle trajectories were extracted using the TrackMate plugin in Fiji (for an average of 20 particles), and subsequent data processing and trajectory analysis were performed using custom MATLAB scripts.

**Results**

To establish a well-defined active colloidal system, we first fabricated asymmetric PS-Pt Janus particles and engineered substrates with controlled topography (Fig. 1). The Pt coating produced a hemispherical metallic cap on the spherical PS colloids (diameter $d_P$ = 2 µm), as confirmed by SEM imaging, yielding active particles capable of catalytic $H_2O_2$ decomposition. In parallel, PDMS replicas of a master pattern generated two distinct topographic patterns: a microscale pattern, OG PDMS with groove height, $h_P$, of 1 µm and width, $l_P$, of 5 µm, and a nanoscale pattern, CD PDMS with groove height, $h_P$, of 120 nm and width, $l_P$, of 750 nm. AFM measurements verified the periodicities and height profiles of both topographies. Together, the combination of catalytic asymmetry and substrate morphology defines a model platform in which internal flows, bubble dynamics, and particle transport can be systematically probed. This framework enables direct comparison of how activity and topography couple during evaporation, motivating the flow and trajectory analyses presented in subsequent figures.

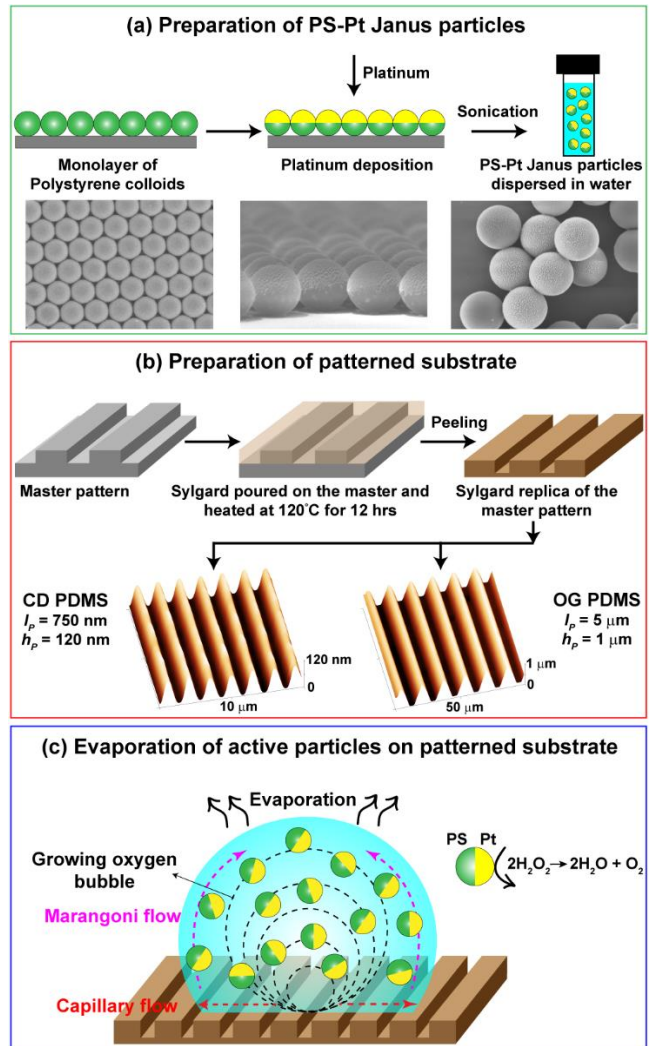

**Fig. 1:** Schematic illustration of the overall workflow, highlighting the (a) preparation of PS-Pt Janus particles, (b) fabrication of patterned PDMS substrates, and (c) evaporative drying of the active particle suspension on a patterned substrate. (a) Monolayer of polystyrene micro-spheres deposited on a substrate, followed by platinum coating and sonication to yield PS-Pt Janus particles dispersed in water. The corresponding SEM images are shown below the schematics. (b) Replication of a master pattern onto PDMS, producing topographically patterned substrates, OG PDMS and CD PDMS. The corresponding AFM images of the PDMS replicas are shown along with the groove width ($l_P$) and height ($h_P$). (c) Drying of active Janus droplets on patterned substrates.

### *Drying kinetics of PS-Pt active particles on OG PDMS.*
The drying of PS-Pt Janus droplets on OG PDMS substrates reveals a systematic transition in evaporation dynamics and deposit morphologies with increasing $H_2O_2$ concentration. Fig. 2 summarizes how increasing $H_2O_2$ concentration (0, 1, 5 and 10 wt% in Milli-Q water) modifies the dried morphologies of active droplets on OG PDMS substrates (Figs. 2a(1)-2d(1)). Corresponding magnified images are shown in Figs. 2a(2)-2d(2). The drying kinetics evolve systematically with $H_2O_2$ concentration (Figs. 2a(3)-2d(3)). Insets show the magnified side-views of the evaporating droplets at different times during the drying process. These changes are mirrored in the intermediate droplet shapes (Figs. 2a(4)-2d(4)); a stable spherical cap at low activity transitions to visibly distorted and bubble-containing droplets as catalytic activity intensifies. The resulting internal flow reorganization is captured in the particle trajectories (Figs. 2a(5)-2d(5)). Corresponding data for $H_2O_2$ concentrations of 0.1, 3 and 7.5 wt% are shown in SI Appendix, Fig. S1. These $H_2O_2$ concentration-dependent transitions converge into four distinct regimes, as shown in Fig. 2e, which shall be discussed in detail in this section.

At 0 wt% $H_2O_2$ (no catalytic activity), evaporation proceeds under the influence of capillary-driven outward flow and contact line pinning. The pinned three-phase contact line sustains a radial flux that transports particles steadily toward the edge, where they accumulate into a ring-like deposit, as observed from the final dried morphology shown in Figs. 2a(1) and 2a(2). Strong pinning of the droplet at the contact line is evident from the fact that the initial diameter of the drop is similar to the final dried deposit diameter (Figs. 2a(1) and 2a(4)). From the contact angle vs time graph shown in Fig. 2a(3), we verify that evaporation proceeds in an almost constant contact radius mode (CCR), resulting in strong pinning and thick ring formation at the droplet edge. Although the substrate grooves can act as potential traps, they are not sufficiently deep to retain a significant fraction of the particles against the dominant capillary flow field. Consequently, most particles accumulate at the contact line, giving rise to a pronounced coffee-ring morphology. At 0.1 wt% $H_2O_2$ (SI Appendix, Fig. S1), the droplet continues to evaporate under a predominantly capillary-driven mode, as indicated by the smooth decay of the contact angle over time. However, unlike the continuous and sharply defined ring obtained at 0 wt%, the final deposit displays localized thinning and some regions of uneven particle accumulation along the periphery (SI Appendix, Fig. S1a(1)), marking the first deviation from classical coffee-ring formation. Since $H_2O_2$ exhibits a higher surface tension ($\gamma$) than pure water (34), even slight fuel enrichment near the edge due to preferential water evaporation leads to $\gamma_{edge} > \gamma_{center}$, establishing a surface tension gradient directed radially outward (6). Here, $\gamma_{edge}$ is the surface tension at the edge, and $\gamma_{center}$ is the surface tension at the bulk (Fig. 2e). As this outward surface motion develops, mass conservation establishes a compensating inward return flow through the bulk, giving rise to intermittent solutal Marangoni pulses that momentarily modulate the outward capillary flux.

On increasing the $H_2O_2$ concentration to 1 wt %, the contact angle evolution (Fig. 2b(3)) no longer follows a smooth monotonic decay but instead displays distinct spikes followed by abrupt drops, which we attribute to oxygen bubble growth and collapse cycles inside the droplet. Each temporary rise in contact angle corresponds to bubble nucleation at the substrate followed by bubble growth, which momentarily increases the droplet curvature. This is immediately followed by a sharp fall associated with bubble bursting (43), which causes a sudden drop in contact angle, and can be correlated to hydrodynamic shock events. The final dried morphology (Fig. 2b(1)) reflects these intermittent instabilities. Instead of a single, continuous coffee-ring, the deposit exhibits fragmented rim segments and inner particle clusters, with localized void-like regions near the perimeter (Fig. 2b(2)), consistent with bubble footprints disrupting particle deposition upon collapse. These features indicate that particle transport no longer occurs through a steady outward flux, but rather through pulsed deposition triggered by bubble-induced radially inward Marangoni flow. As reported for catalytic bubble systems, growing oxygen bubbles establish localized Marangoni circulations at their perimeter, which draw fluid inward, counteracting capillary flow (34). Similarly, at 3 wt% $H_2O_2$, bubble nucleation becomes frequent and sustained throughout evaporation (SI Appendix, Fig. S1b).

At 5 wt% $H_2O_2$, the droplet interior becomes populated by multiple bubbles that nucleate, coalesce, and persist for extended periods. The contact angle curve (Fig. 2c(3)) shows frequent oscillations of smaller amplitude, reflecting quasi-steady cycles of bubble expansion, coalescence, and partial collapse. The intermediate drying image (Fig. 2c(4)) reveals that several bubbles anchor at the grooves, demonstrating strong coupling between substrate topography and catalytic gas generation. These transiently pinned bubbles are locally immobilized on the interface and force evaporation to proceed non-uniformly across the droplet footprint. The resulting deposition is discontinuous, with clustered particle aggregates surrounding bubble pinning sites and patchy, low-density regions elsewhere (Fig. 2c(1)).

At 10 wt % $H_2O_2$, catalytic decomposition nears saturation, resulting in rapid and sustained oxygen generation throughout the droplet. The contact angle curve (Fig. 2d(3)) exhibits large amplitude oscillations and abrupt discontinuities, indicating repeated bubble nucleation, expansion, and explosive collapse. The droplet surface becomes highly irregular, with bubbles covering most of the footprint (Fig. 2d(4)). The resulting deposition (Fig. 2d(1)) is extremely heterogeneous, composed of large circular voids (former bubble footprints) surrounded by particle-rich regions. This represents the bubble vortex-dominated regime, where bubble dynamics fully override capillary and Marangoni control. Successive bursts produce strong internal flows that prevent the formation of any coherent ring or film. At this stage, the evaporation process is characterized by rapid pressure fluctuations, chaotic recirculation, and non-repeatable deposit topology, which represent the extreme limit of catalytic drying behavior. Similar behavior is also observed at $H_2O_2$ concentration of 7.5 wt% (Fig. S1c).

Also, particle trajectories reveal a clear progression in internal flows and transport directions as $H_2O_2$ concentration increases (Fig. 2a(5)-d(5)). In the absence of fuel (0 wt% $H_2O_2$, Fig. 2a(5)), particles move radially outward in straight paths toward the pinned contact line under capillary-driven advection. The color-coded trajectories indicate outward flow. We estimated low and nearly constant speeds throughout evaporation (Fig. 2e). At 0.1 wt% $H_2O_2$ (SI Appendix, Fig. S1a5), we observe small fluctuations indicating unsteady, oscillatory transport, yet the dominant direction remains outward. At 1 wt% $H_2O_2$ (Fig. 2b(5)), the first bubble nucleation events occur, initiating flow reversals and trajectories showing persistent inward motion. On increasing the $H_2O_2$ concentration to 5 wt% (Fig. 2c(5)), bubbles become stable and remain pinned to the substrate. Particles near the droplet interface experience accelerated motion as Marangoni stresses draw fluid toward the periphery of the bubble. At 10 wt% $H_2O_2$, multiple bubbles nucleate, grow, coalesce, and burst. Trajectories exhibit frequent crossings, reversals, and sharp angular turns, characteristic of chaotic advection rather than coherent flow (Fig. 2d(5)). Particles are first drawn toward the bubble perimeter (vortex capture), and then are violently expelled upon bubble collapse. Such bubble rise and collapse have also been observed on flat surfaces (44,45). The propulsion speed versus $H_2O_2$ concentration analysis (Fig. 2e) quantifies this trend; average particle speed increases nearly linearly with $H_2O_2$ concentration, reflecting increased catalytic turnover and stronger interfacial Marangoni stresses. The schematic in Fig. 2e visualizes how the direction of internal flow evolves with activity.

The schematic in Fig. 2f provides a visual overview of the drying process on OG PDMS substrates with pattern length scales commensurable with particle size, and highlights how multiple flow modes co-exist and evolve during evaporation. Overall active droplet evaporation on OG gratings, which act as transient pinning sites for bubbles, demonstrates a clear progression from capillary-dominated coffee-ring formation (0 wt% $H_2O_2$) to solutal Marangoni-influenced mixed transport (0.1-1 wt % $H_2O_2$), to bubble-dominated inward Marangoni flow (1-5 wt % $H_2O_2$), and finally to a bubble vortex-driven flow (7.5-10 wt % $H_2O_2$). This coupling between catalytic activity, bubble dynamics, and substrate topography defines the final drying patterns of active colloidal droplets.

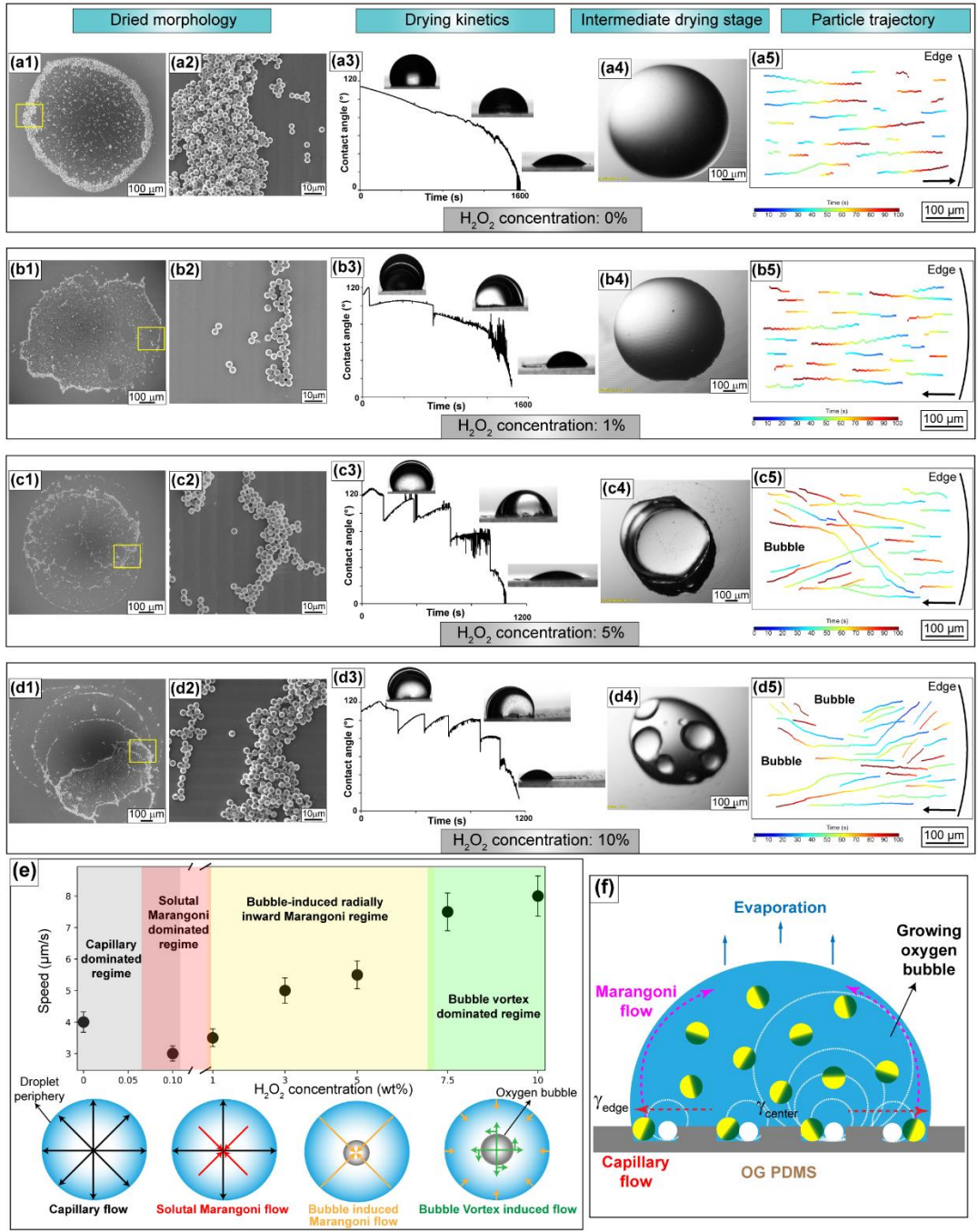

**Fig. 2:** Morphological variations in the final dried droplets and drying kinetics on OG patterned hydrophobic PDMS replicas with increasing $H_2O_2$ concentrations, 0, 1, 5, and 10 wt% respectively. (a1-d1) SEM images of final dried morphologies at selected fuel concentrations, (a2-d2) high magnification images of transitions from ring-like deposits to patchy and porous aggregates. (a3-d3) Corresponding contact angle versus time plots showing the drying modes, with representative droplet profiles at different stages. The mismatch in the illumination is an artefact of side-view imaging. (a4-d4) Optical micrographs of top-views of droplets during intermediate stages of drying. (a5-d5) Particle trajectories corresponding to each case with color maps shown below. (e) Top: Propulsion speed versus $H_2O_2$ concentration. Colored regions show the overlapping four dominant flow regimes. Bottom: schematic illustration of the flows at different activities. (f) Schematic illustration of bubble activity during droplet drying on OG PDMS substrates.

***Drying kinetics of PS-Pt active particles on CD PDMS.***
While OG substrates are able to confine particles and locally enhance the catalytic turnover, CD substrates, because of their nanoscale groove dimensions, cannot physically confine or guide individual particles. This geometric mismatch alters how catalytic activity, Marangoni stresses, and gas generation couple to evaporation. Fig. 3 summarizes how increasing $H_2O_2$ concentration (0, 1, 5 and 10 wt%) modifies the drying front of the active droplets on the CD PDMS (Fig. 3a(1)-3d(1)). The final dried deposit clearly shows a signature of this asymmetric drying and particle chains (Fig. 3a(2)-3d(2)). Furthermore, the drying kinetics evolve systematically with $H_2O_2$ concentration (Fig. 3a(3)-3d(3)). The resulting internal flow reorganization is captured in the particle trajectories (Fig. 3a4-3d4). Corresponding data for $H_2O_2$ concentration of 0.1, 3 and 7.5 wt% are shown in SI Appendix, Fig. S2. These $H_2O_2$ concentration dependent transitions converge into three distinct regimes, as shown in Fig. 3e, which shall be discussed in detail in this section.

At 0 wt% $H_2O_2$, evaporation on CD PDMS is governed entirely by outward capillary flow. The droplet shows stick-slip motion as the contact line intermittently pins and depins at the grooves (Fig. 3a(3)), but the overall evaporation proceeds through a predominantly CCR mode. The final deposit displays a thick coffee-ring formation (Fig. 3a(2)) and lacks any groove localized accumulation, confirming that the CD topography does not significantly perturb the capillary dominated transport pathway. On addition of slight activity to the system (SI appendix Fig. S2 for 0.1 wt% $H_2O_2$), outward capillary flow still dominates, but preferential water evaporation leads to slight $H_2O_2$ enrichment at the droplet edge. Because $H_2O_2$ exhibits higher surface tension than water, this enrichment establishes a weak outward interfacial gradient, generating intermittent solutal Marangoni pulses that mildly oppose the capillary flux. The morphology remains nearly uniform, but subtle thinning near parts of the periphery indicates the earliest deviations from classical coffee-ring behavior on CD PDMS.

At 1 wt% $H_2O_2$, solutal Marangoni stresses intensify and begin to counteract the outward capillary flux. However, unlike on the OG substrate, where the grooves enhance oxygen retention and triggers bubble formation and growth, the nanoscale grooves of the CD substrate disperse the Janus particles. The reduced catalytic activity at the substrate prevents localized oxygen supersaturation and suppresses bubble nucleation and growth. Consequently, no visible bubbles are observed at this concentration cand the grooves remain oxygen rich guiding the drying front along the lateral direction (Fig. 3b(1)). The dried morphology becomes more uniform across the interior (Fig. 3b(2)), with reduced edge accumulation and elongated patterns aligned with the CD topography. At 3 wt% $H_2O_2$ (SI appendix Fig. S2b), solutal Marangoni stresses strengthen sufficiently to exceed the outward capillary flux, reversing the net transport direction toward the droplet interior. Despite the increased catalytic activity, the nanoscale grooves prevent localized oxygen buildup, and no visible bubble nucleation occurs. This sharply contrasts with OG PDMS, where microgrooves enhance oxygen retention and trigger early bubble formation at the same $H_2O_2$ concentration. The receding drying front becomes more strongly guided along the groove orientation, and particle deposition begins to adopt anisotropic, groove parallel alignment (SI Appendix, Fig. S2b(2)). These elongated interior aggregates indicate that topographic guidance rather than bubble-mediated flow dominates particle placement at this stage.

At 5 wt% $H_2O_2$ bubble activity begins to emerge. The contact angle evolution exhibits pronounced fluctuations (Fig. 3c(3)), and the drying videos show small, rapid flickering of the droplet interface, both consistent with bubble nucleation and collapse events occurring below the optical detection threshold. These microbursts introduce intermittent disturbances to the interface and initiate bubble-induced inward Marangoni flow. However, because the nanoscale grooves disperse catalytic activity and prevent sustained oxygen supersaturation, these bubbles remain transient and never grow into stable bubbles as observed on OG PDMS. The drying front continues to follow the groove direction, and particles assemble into elongated, groove parallel aggregates (Fig. 3c(1) and 3c(2)), indicating that topographic guidance still dominates, even as early-stage bubble driven stresses begin to modulate the transport field.

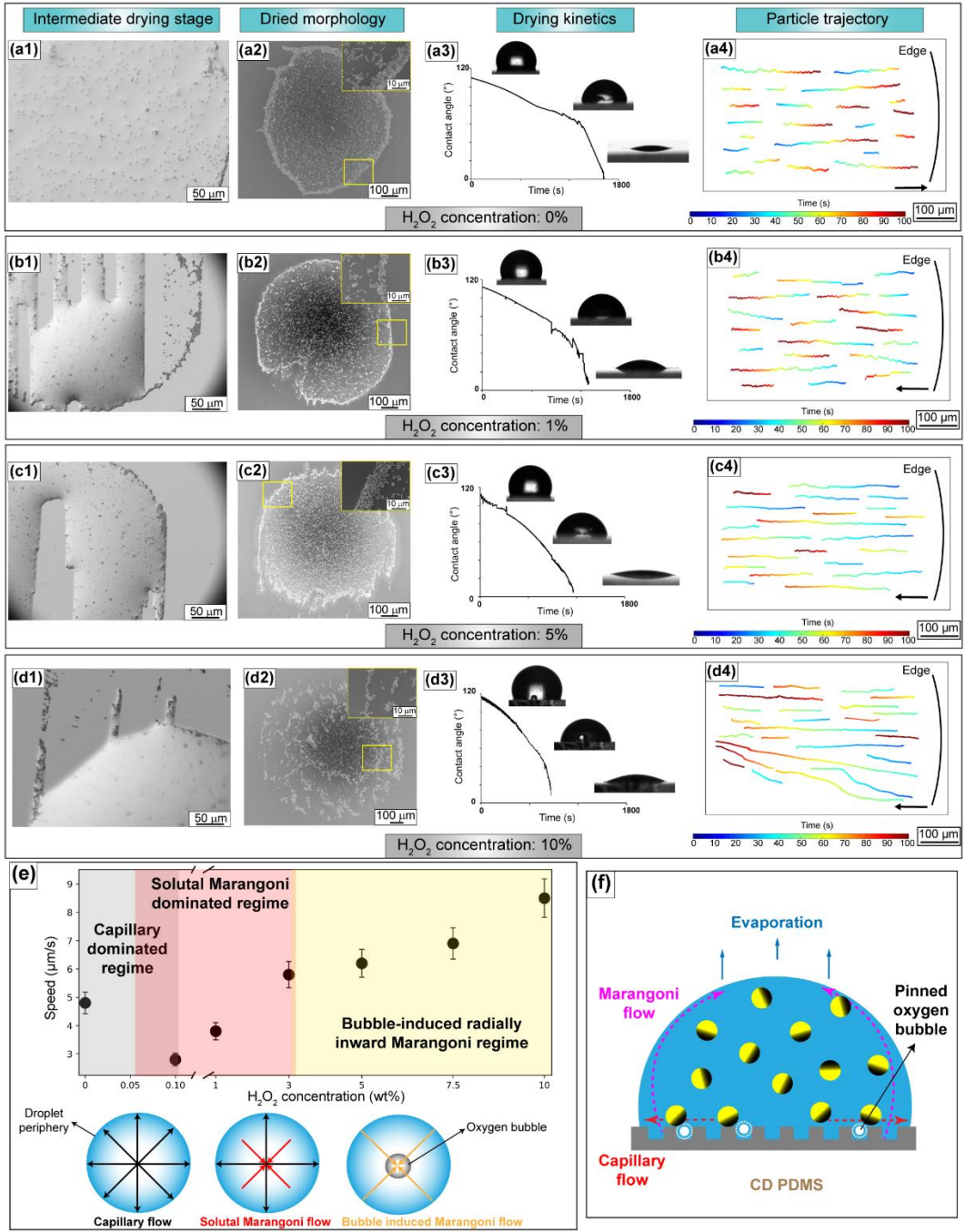

**Fig. 3:** Morphological variations in the final dried droplets and drying kinetics on CD patterned hydrophobic PDMS replicas with increasing $H_2O_2$ concentrations, 0, 1, 5, and 10 wt% respectively. (a1-d1) Optical micrographs of top-views of droplets during intermediate stages of drying. (a2-d2) SEM images of final dried morphologies at the same fuel concentrations. Inset: high magnification images of transitions from ring-like deposits to patchy and porous aggregates. (a3-d3) Corresponding contact angle versus time plots showing the drying modes, with representative droplet profiles at different stages. (a4-d4) Particle trajectories corresponding to each case with color maps shown below. (e) Top: Propulsion speed versus $H_2O_2$ concentration. Colored regions show the overlapping three dominant flow regimes. Bottom: schematic illustration of the flows at different activities. Here topographic guidance overpowers the bubble dominated flow. (f) Schematic illustration of bubble activity for high $H_2O_2$ concentration during droplet drying on CD PDMS substrates.

At higher concentrations (7.5-10 wt% $H_2O_2$), catalytic oxygen production becomes sufficiently intense to enable bubble nucleation. However, unlike on OG substrates where bubbles grow, rise, and induce strong vortex-like flows, the bubbles on CD PDMS remain pinned to the surface. The combination of nanoscale roughness and shallow groove geometry prevents bubble detachment and suppresses vertical migration. As a result, bubble dynamics are subdued, pinned bubbles do not generate toroidal circulations, and no vortex-driven recirculation is observed. Instead, bubbles act as local obstacles that redirect particle motion and promote the formation of chain-like aggregates aligned along the groove orientation (Fig. 3d(2)).

In conclusion, CD patterned PDMS establishes a unique evaporation regime. While its nanoscale grooves cannot confine individual particles, they strongly regulate the macroscopic drying front and effectively suppress bubble-mediated flows across a wide range of catalytic activity. Bubbles appear only at the highest $H_2O_2$ concentrations, though they remain pinned and fail to generate the vigorous vortex flows characteristic of OG substrates. Consequently, the final morphologies on CD PDMS are dominated not by bubble-driven disorder but by elliptical, groove aligned assemblies. This contrast with the OG substrate underscores how geometric confinement at the particle-substrate interface governs oxygen retention, bubble dynamics, and the resulting circular deposition pathways.

Particle trajectories on CD PDMS (Fig. 3a(4)-d(4)) reinforce this picture. At 0 wt% $H_2O_2$, particles follow outward radial paths under capillary driven transport, with negligible resistance from the shallow grooves. At 1 wt%, the emergence of inward solutal Marangoni stresses reverses particle motion toward the droplet center, consistent with the suppression of edge accumulation and the absence of bubbles. At 10 wt%, bubble nucleation finally occurs, but the resulting bubbles remain surface pinned and generate only local perturbations, unlike the strong vortex flows observed on OG PDMS. Although particle speed increases with $H_2O_2$ concentration (Fig. 3e) reflecting enhanced catalytic turnover, the trajectories are smoother and due to lack of vortex induced accelerations. The schematic in Fig. 3f provides an integrated overview of the drying process on CD PDMS, illustrating how flow modes evolve when nanoscale grooves are unable to confine individual particles. Evaporation on CD gratings progresses from a capillary dominated regime (0 wt% $H_2O_2$), to a solutal Marangoni modulated transport regime (0.1-3 wt% $H_2O_2$), and then to a transitional regime where subcritical bubble activity begins but remains weak (5 wt%). Only at high concentrations (7.5-10 wt% $H_2O_2$) do bubbles nucleate, yet the nanoscale groove roughness pins them to the substrate and prevents vertical growth, suppressing vortex driven flows.

***Drying dynamics of active Janus particle droplets on superhydrophilic OG PDMS.***
In order to analyze bubble activity in more detail, we studied the drying of active droplets on OG (Philic) patterned substrates. Fig. 4 summarizes how increasing $H_2O_2$ concentration (0, 5 and 10 wt%) modifies the dried deposit of active droplets on OG PDMS (Philic) (Fig. 4a(1)-4c(1)), with magnified images shown in Fig. 4a2-4c2. The side-view and top-view images of the droplet at 0 wt% $H_2O_2$ concentration are shown in Fig. 4d(1) and 4d(2). Corresponding data for $H_2O_2$ concentrations of 0.1, 1, 3 and 7.5 wt% are shown in SI Appendix, Fig. S3. The schematic in Fig. 4e provides a visual overview of the drying process on OG PDMS (Philic) substrates, and highlights how multiple flow modes co-exist and evolve during evaporation.

On plasma treated OG substrates (initial contact angle ~ 10°), the dispensed droplet spreads into a high aspect ratio ultrathin film (thickness only a few micrometers, Fig. 4d). This geometry eliminates the vertical height gradients that normally sustain strong capillary flow in sessile droplets. This suppresses the coffee-ring effect and results in a uniform deposition of particles throughout the droplet periphery (Fig. 4a). Evaporation becomes nearly spatially uniform, and particle motion is governed by interfacial solutal gradients in the presence of $H_2O_2$. Immediately after spreading, the particles are immobilized at the contact line. Unlike the hydrophobic substrate where evaporative flux at the rim enriches $H_2O_2$, here the edge becomes a localized catalytic sink. Pt hemispheres at the perimeter continuously decompose $H_2O_2$, and since the film is extremely thin, diffusion of fresh $H_2O_2$ from the droplet interior is slower than its catalytic consumption at the edge. This establishes a lateral outward $H_2O_2$ concentration gradient. Since surface tension increases with local $H_2O_2$ concentration, this corresponds to: $\gamma_{center} > \gamma_{edge}$. As Marangoni stresses drive interfacial flow from the low $\gamma$ at the edge toward the high $\gamma$ at the center; conservation of mass forces a compensating bulk flow outward, which transports Janus particles from the center to the edge. Thus, although the surface flow is inward, the observable particle motion is outward.

Unlike the hydrophobic case, where evaporation drives $H_2O_2$ enrichment at the edge and produces inward surface flow, plasma treated surfaces invert the balance of processes; catalytic depletion dominates over evaporation, flipping the gradient and therefore also the Marangoni flow direction. As a result, a persistent outward trajectory at every $H_2O_2$ concentration (0.1-10 wt %) is observed. Also, dissolved $O_2$ rapidly vents out of the thin film, preventing bubble growth or pinning. Thus, no vortex trapping or bubble growth events occur, which is in stark contrast to bubble-induced chaotic flows on hydrophobic OG PDMS. At higher peroxide concentrations (7.5-10 wt %), the edge becomes increasingly fuel depleted, reducing self-phoretic repulsion (46) and allowing particles to assemble into dense peripheral aggregates (Fig. 4c). The combination of sustained outward advective flux and weakened phoretic repulsion produces tight clustering exclusively at the rim, even though no bubbles are present. Thus, substrate wettability not only changes droplet geometry, it reverses the underlying Marangoni mechanism. This reveals that wettability controls the sign of the solutal Marangoni stress, the direction of particle transport and eventually, the deposit morphologies after the completion of droplet drying.

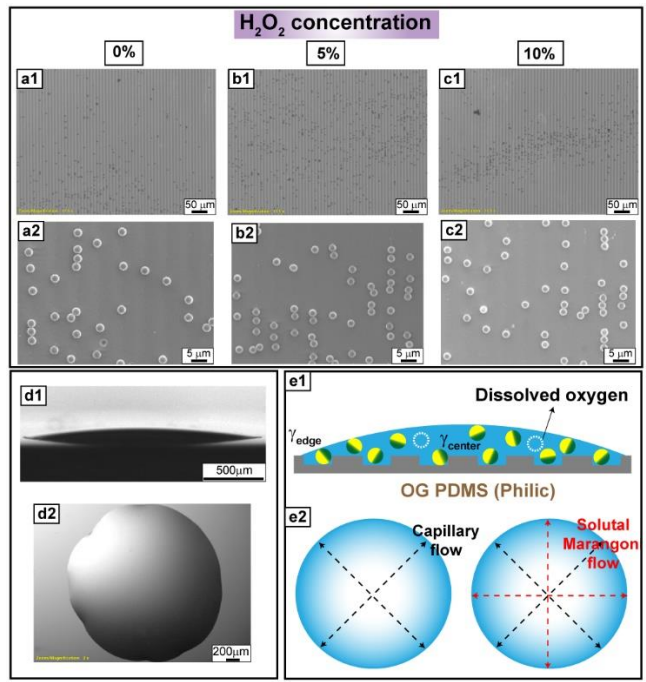

**Fig. 4:** Evaporation of Janus particle-laden droplets on plasma-treated OG patterned PDMS (Philic) substrates (initial contact angle ~10°) as a function of $H_2O_2$ concentration. (a1-c1) Final deposition patterns for 0, 5 and 10 wt % $H_2O_2$, observed under an optical microscope, (a2-c2) magnified particle images observed under SEM. (d) Side-view and top-view images of the droplet immediately after dispensing (corresponding to (a)), confirming rapid spreading into a thin film. (e) Schematic illustration of flow direction during droplet drying o OG PDMS (Philic) substrate.

***Mechanistic framework of flow regimes during fuel-driven droplet evaporation.***

The schematic shown in Fig. 2e consolidates the particle trajectory data into a mechanistic framework describing how internal flow on OG PDMS substrates evolves with increasing catalytic activity. Control experiments with passive PS particles in $H_2O_2$ (no Pt) show exclusively radial outward motion, identical to the classical coffee-ring flow on OG PDMS substrate. In these droplets, evaporation drives a capillary flux toward the pinned contact line. Although water evaporates faster than $H_2O_2$, creating a compositional difference, the lateral concentration difference $\Delta c$ along the air-liquid interface remains small because there is no catalytic consumption of $H_2O_2$. The strength of Marangoni circulation is quantified using the solutal Marangoni number (6,18): $Ma = \frac{(\partial\gamma/\partial c)\,\Delta c\, L}{\mu\, D}$, where $\partial\gamma/\partial c$ is the change of surface tension with solute concentration, L is the lateral length scale of droplet, $\mu$ is the dynamic viscosity, and D is the diffusivity of $H_2O_2$. In passive droplets, diffusion of $H_2O_2$ is fast, and $Ma \ll 1$. With no surface tension gradient to apply tangential stress, no recirculation loop forms, and the flow remains outward everywhere in the droplet.

In contrast, when PS-Pt Janus particles are dispersed in $H_2O_2$, catalytic decomposition locally removes $H_2O_2$ at the Pt cap and simultaneously produces dissolved $O_2$. Together with preferential evaporation of water at the contact line, this creates a lateral surface tension gradient $\tau_M$ along the droplet interface, which is given by the equation (6,18): $\tau_M = \nabla_S \gamma = (\partial\gamma/\partial c)\nabla_S c$, where $\tau_M$ is the Marangoni shear stress acting along the interface, $\nabla_S$ is the surface (tangential) gradient operator and *c* is interfacial concentration gradient of $H_2O_2$. Because $H_2O_2$ has higher surface tension than water, enrichment of $H_2O_2$ at the droplet edge produces $\gamma_{edge} > \gamma_{center}$. Fluid always flows from low $\gamma$ to high $\gamma$, therefore, Marangoni shear drives outward surface flow, and conservation of mass forces a compensating inward return flow in the bulk. In our observations, focused on the bulk, only the inward leg of this circulation cell is visible, appearing as the outward to inward flow reversal. At higher $H_2O_2$ concentrations (≥ 3 wt %), catalytic turnover is fast enough to nucleate $O_2$ bubbles. Bubble growth amplifies surface tension gradients and creates local recirculation zones, and bubble collapse produces impulsive flow bursts. This generates vortex captured trajectories, direction switching, and chaotic advection. In this limit, $\Delta c$ becomes highly localized around bubbles, and $Ma \gg 1$, such that Marangoni and bubble-driven interfacial stresses dominate over capillary flow.

**Conclusion**

The comparative studies on OG and CD patterned PDMS replicas reveal that groove dimension critically dictates evaporation dynamics and particle transport. On OG surfaces, the micron scale grooves confine Janus particles effectively, promote localized catalytic activity, and facilitate bubble nucleation even at low $H_2O_2$ concentrations, with bubbles growing and rising freely. In contrast, the nanoscale grooves of CD substrates do not confine the particles; oxygen generation remains dispersed, and bubbles only appear at very high $H_2O_2$ concentrations, where they remain pinned and unable to rise. The nanoscale texture of CD also induces elliptical droplet footprints, with preferential spreading along the groove axis, whereas OG droplets retain circular symmetry. At high fuel concentrations, CD substrates further promote aggregated deposits aligned along the grooves. Overall, these findings establish that groove dimension governs not only bubble nucleation and motion but also anisotropic wetting and final deposit morphology, offering a tunable design principle for controlling active droplet evaporation.

**Supporting Information for**

Bubble-Driven Flow Transitions in Evaporating Active Droplets on Structured Surfaces.

Meneka Banik and Ranjini Bandyopadhyay*

*Corresponding author

Email: ranjini@rri.res.in

**Figures**

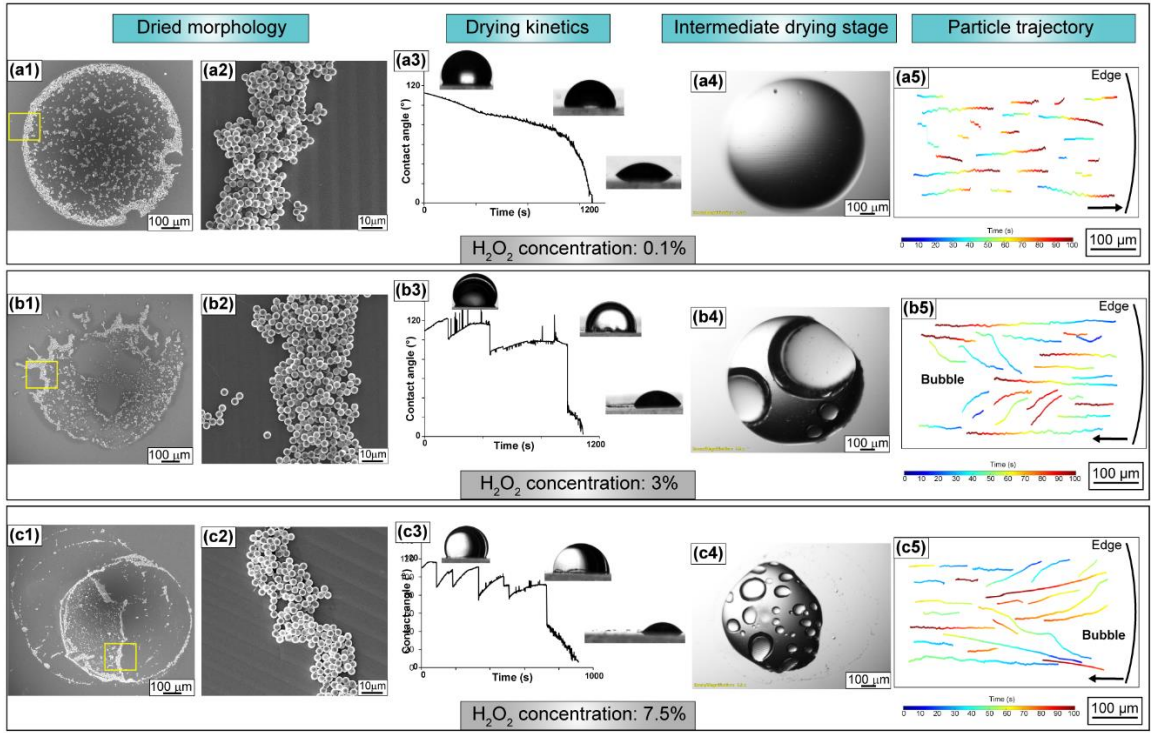

**Fig. S1.** Morphological variations in the final dried droplets and drying kinetics on OG patterned hydrophobic PDMS replicas with increasing $H_2O_2$ concentrations, 0.1, 3 and 7.5 wt% respectively. (a1-c1) SEM images of final dried morphologies at selected fuel concentrations, (a2-c2) high magnification images of transitions from ring-like deposits to patchy and porous aggregates. (a3-c3) Corresponding contact angle versus time plots showing the drying modes, with representative droplet profiles at different stages. (a4-c4) Optical micrographs of top-views of droplets during intermediate stages of drying. (a5-c5) Particle trajectories corresponding to each case with color maps shown below.

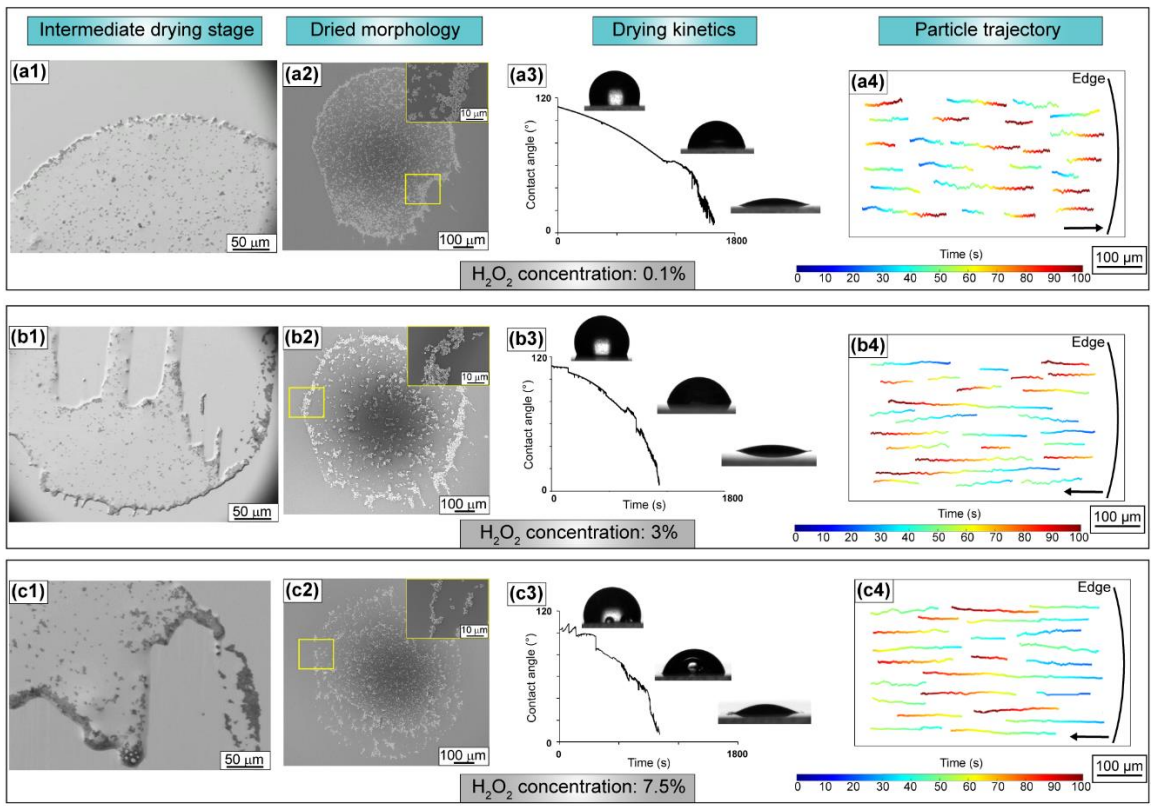

**Fig. S2.** Morphological variations in the final dried droplets and drying kinetics on CD patterned hydrophobic PDMS replicas with increasing $H_2O_2$ concentrations, 0.1, 3 and 7.5 wt% respectively. (a1-c1) Optical micrographs of top-views of droplets during intermediate stages of drying. (a2-c2) SEM images of final dried morphologies at selected fuel concentrations. Inset: high magnification images of transitions from ring-like deposits to patchy and porous aggregates. (a3-c3) Corresponding contact angle versus time plots showing the drying modes, with representative droplet profiles at different stages. (a4-c4) Particle trajectories corresponding to each case with color maps shown below.

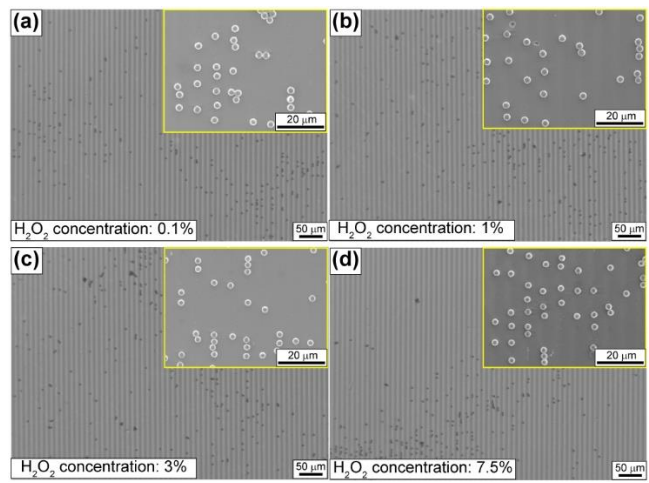

**Fig. S3.** Morphological variations in the final dried droplets on OG patterned hydrophilic PDMS replicas with increasing $H_2O_2$ concentrations, 0.1, 1, 3 and 7.5 wt% respectively (a-d). Corresponding insets are magnified images captured at the center of the deposits.